\documentstyle[12pt,epsf]{article}
\newcommand{\la}{\langle}
\newcommand{\ra}{\rangle}

\newcommand{\beq}{\begin{eqnarray}}
\newcommand{\eeq}{\end{eqnarray}}
\newcommand{\Sigpi}{\Sigma _{\pi N}}

\newcommand{\btem}{\bibitem}

\begin{document}
\title{ 
SIGNIFICANCE OF THE SIGMA MESON IN HADRON PHYSICS (QCD) AND
 POSSIBLE EXPERIMENTS TO OBSERVE IT
\footnote{Invited talk presented at the Workshop on Hadron Spectroscopy
(whs99), Frascati- INFN, March 8 - 12, 1999.}}
\author{
Teiji Kunihiro        \\
{\em Faculty of Science and Technology, Ryukoku University},\\
{\em  Seta, Otsu, 520-2194, Japan} \\
}
\maketitle
\baselineskip=14.5pt
\begin{abstract}
We first discuss the theoretical and phenomenological
significance of the sigma meson ($\sigma$) in QCD.
It is indicated that if the collective modes with the mass 500-600 MeV
exists in the $I=J=0$ channel, various empirical facts in hadron physics 
can be naturally accounted for, which  otherwise would remain mysterious.
We propose several experiments to produce and detect the $\sigma$
 in nuclei using nuclear and electro-magnetic projectiles. 
The recent CHAOS data which show
a spectral enhancement near the 2 $m_{\pi}$ threshold in the $\sigma$
channel from the reactions A$(\pi, 2\pi)$A' where 
A and A' denotes nuclei 
is interpreted as a possible evidence of a partial
restoration of chiral symmetry in nuclei.
\end{abstract}
\baselineskip=17pt
\section{Introduction}
The sigma ($\sigma$) 
meson is the chiral partner of the pion for the 
$SU_L(2)\otimes SU_R(2)$ chiral symmetry in QCD.
The   particle representing the quantum fluctuation of the order parameter
 $\tilde {\sigma}\sim \la (:\bar q q:)^2\ra$ is the $\sigma$ meson. 
The $\sigma$ meson is analogous
 to the Higgs particle in the Weinberg-Salam theory.
Some effective theories\cite{HK94} including the ladder QCD\cite{scadron}
 and Weinberg's mended symmetry\cite{weinberg} 
predict the $\sigma$ meson mass $m_{\sigma}\sim 500$-700 MeV:
The Nambu-Jona-Lasinio(NJL) model\cite{njl}
 is now widely used as an effective theory 
for describing the chiral properties of QCD\cite{klevansky,HK94}.
In this model, the chiral symmetry is realized linearly like the linear
sigma models, hence the appearance of the $\sigma$ meson is 
inevitable; one has the $\sigma$ meson as well as the pions, and 
 the chiral symmetry makes $m_{\sigma}$ 
 twice of the constituent quark mass $M_q\sim 335$ MeV as well as 
 the pions are massless in the chiral limit
 \cite{njl},  hence 
\beq
m_{\sigma}\sim 2M_q\sim 670 {\rm MeV}.
\eeq
The significance of this relation in the context of QCD was 
emphasized by us in \cite{ptp85}.
If  such a  scalar meson with a low mass is identified, many 
 experimental facts  which otherwise are mysterious  can be nicely accounted 
for  in a simple way\cite{HK94,ptp95}: The correlation in the scalar
 channel as summarized by such a scalar meson can account for the 
 $\Delta I= 1/2$ rule for  the decay process
 K$^{0} \rightarrow \pi^{+}\pi^{-}$ or 
 $\pi^{0}\pi^{0}$ \cite{morozumi}.
In the meson-theoretical model for the nuclear
 force,   a scalar meson exchange with the mass range  500$\sim $ 700 MeV 
is  indispensable to fully account for the state-independent attraction in 
the  intermediate range.
The collective excitation in the scalar channel
  as described as   the $\sigma$ meson is essential 
 in reproducing the empirical value of the $\pi$-N sigma term 
 $\Sigpi =\hat {m}\la \bar{u} u + \bar {d} d\ra$\cite{kuni88},
the empirical value of which is reported to be $45\pm 10$ MeV,
while the naive chiral perturbation theory fails in reproducing the empirical 
value unless an unrealistically  large strangeness content of the proton is 
assumed.
 We remark also the convergence radius of the chiral perturbation theory
\cite{CH}  is
 linked with the mass of the scalar meson.  
 
  These facts indicate that the scalar-scalar
 correlation is important in the hadron dynamics. This is in a sense
 natural because the dynamics which is responsible for the correlations
 in the scalar channel is nothing but the one which drives the chiral
 symmetry breaking.\cite{ptp95}  
  
A tricky point on the $\sigma$ meson is that the $\sigma$ meson strongly couples 
to two pions which  gives rise to a large width $\Gamma \sim m_{\sigma}$.
Recent phase shift analyses of the pi-pi scattering in the scalar channel
 claim a pole of  the scattering matrix in the complex energy plane  
 with the real part Re$m_{\sigma}= 500$-700 MeV and the imaginary part
 Im$m_{\sigma}\simeq 500$MeV\cite{pipi}, although
the possible coupling with glue balls with $J^{PC}= 0^{++}$ make the 
 situation obscure.
Our view about the identification of the scalar mesons is given in
 chapter 3 of ref.\cite{HK94}.


\section{Experiments to produce the $\sigma$ using nuclear targets}

We  have seen that the correlations which may be summarized by the 
 unstable and hence elusive $\sigma$ meson play significant roles in the
 hadron phenomenology at low energies.  Therefore one may wonder
 whether  there is any chance   to  observe the $\sigma$ meson clearly. 
 What does come when the environment is changed by rising  temperature
 and/or density? 
 As was  first shown by us\cite{HK84,HK94}, 
 the $\sigma$  decreases
 the mass (softening) 
in association with the chiral restoration in the hot and/or 
dense medium, and  the width of the meson is also expected to decrease
 because the pion hardly changes the mass as long as the system is in 
 the Nambu-Goldstone phase. 
 Thus one can  expect a  chance to see the $\sigma$ meson as a sharp 
 resonance at high temperature and/or density.
Such a behavior of the meson may be detected by observing two pion with 
 the invariant mass around several hundred MeV in relativistic heavy 
ion collisions. 
When the charged pions have 
 finite chemical potentials, the process $\sigma \rightarrow \gamma
 \rightarrow $2 leptons can be used to detect the $\sigma$ meson.

 It is worth mentioning that 
 the simulations of the lattice QCD \cite{lattice} show the decrease of the 
 screening 
mass $m^{\sigma }_{\rm scr}$ of the $\sigma$ meson. Here a screening mass is
 defined through the space correlation of the relevant operator rather 
 than the time correlation as a dynamical (real) mass. 
The relation between  the screening mass and the dynamical mass as 
discussed in \cite{HK84} is not clearly understood yet. 
 Nevertheless it is known \cite{FRIMAN} that the NJL model 
gives the similar behavior for the screening masses in the scalar 
channels with the dynamical ones. It means that the lattice result
 on the screening masses may suggest that the dynamical masses also 
 behave in a way as predicted in \cite{HK84}.
  
Some years ago, the present author
 proposed several experiments \cite{ptp95}
to possibly produce the $\sigma$ meson in nuclei, thereby 
have a clearer evidence of  the  existence of 
the $\sigma$ meson and also explore the possible 
restoration of chiral  symmetry in the nuclear medium.

The first  reaction is 
 A\ ($\pi$, $\sigma$  N)\ A$'$:
 In this reaction, the charged pion ($\pi ^{\pm}$) is absorbed by a nucleon in
 the nucleus, then the nucleon emits the $\sigma$ meson, which decays into 
 two pions.  To make a veto for the two pions from the rho meson, the produced
 pions should be neutral ones which may be detected through
four $\gamma$'s\cite{4gamma}.
The second reaction is  A\ (N, $\sigma$\  N)\ A$'$;
N may be a proton, 
 deuteron or $^3$He, namely any nuclear projectile, which
 collides with a nucleon in the nucleus, then the
 incident particle will emit the $\sigma$ meson, which decays into two pions.
 One may detect 4 $\gamma$ 's from 2 $\pi ^0$ which is the decay product of 
 the $\sigma$.  The 
 collision  with a nucleon may occur after  the emission of the $\sigma$ meson;
 the  collision process is needed for the energy-momentum matching.
In the detection, one may use the two leptons from the process.
This is possible when the sigma has a finite three 
because of 
the scalar-vector mixing in the 
 system with   a finite baryonic density\cite{WALECKA}.
 This detection may gives a clean data, but the yield might be small.
The third one is
 the reaction which uses $\gamma$ rays.
 The $\gamma$ ray emitted from the electron is  converted to the omega
 meson in accord with the vector meson dominance principle, if the 
 particle has a finite three momentum. The omega meson
 may decay into the $\sigma$ meson in the baryonic medium via the process
 $\omega \rightarrow $ N $\bar{\rm N} \rightarrow \sigma$. The $\sigma$ will
 decay into two pions.

When a hadrons is put in a nucleus,
the hadron will couple strongly to various excitations in the system, 
such as nuclear particle-hole (p-h)  and  $\Delta$-hole excitations, 
simultaneous excitations of them and mesons and so on.
In general, the hadron may dissociate into complicated
 excitation to loose its 
identity in the nuclear medium. 
The relevant quantity is the response function or spectral function 
of the system when the quantum numbers of the hadron are put in.
 A response function in the  energy-momentum
space is essentially the spectral function in the meson channel.
If the coupling of the hadron  with the environment is relatively small,
then there may remain a peak with a small width in the spectral function, 
which  correspond to the hadron; such a peak 
may be viewed as an elementary excitation or a quasi particle known in 
Landau's Fermi liquid theory for fermions. 
It is a difficult problem whether a many-body system can be
 treated as an aggregate of elementary excitations or quasi-particles 
 interacting weakly with each other. 
Then how will the decrease of $m_{\sigma}$ in the nuclear medium\cite{HK84}
 reflect in the spectral function in the sigma channel?

A calculation of the spectral function in the $\sigma$ channel
at finite $T$ 
has been performed with the $\sigma$-2$\pi$ coupling incorporated 
in the linear $\sigma$ model\cite{CH98}; it was shown that 
the enhancement of the spectral function in the $\sigma$-channel
just above the
two-pion threshold can be a signal of the decrease of $m_{\sigma}$, 
i.e., a softening.
Recently, it has been shown \cite{HKS} that the spectral enhancement
associated with the partial chiral restoration  
takes place also at finite baryon density close to 
$\rho_0 = 0.17 {\rm fm}^{-3}$.
Referring to \cite{HKS} for the  explicit model-calculation, let us
 describe the general
 features of the spectral enhancement near the two-pion threshold.
 Consider the propagator 
 of the $\sigma$-meson at rest in the medium :
$D^{-1}_{\sigma} (\omega)= \omega^2 - m_{\sigma}^2 - $
$\Sigma_{\sigma}(\omega;\rho)$,
where $m_{\sigma}$ is the mass of $\sigma$ in the tree-level, and
$\Sigma_{\sigma}(\omega;\rho)$ is 
the loop corrections
 in the vacuum as well as in the medium.
 The corresponding spectral function is given by 
\beq
\rho_{\sigma}(\omega) = - \pi^{-1} {\rm Im} D_{\sigma}(\omega).
\eeq
Near the two-pion threshold, 
${\rm Im} \Sigma_{\sigma}$
$\propto \theta(\omega - 2 m_{\pi}) \sqrt{1 - {4m_{\pi}^2 \over \omega^2}}$
 in the one-loop order.
 On the other hand, partial restoration of chiral
 symmetry indicates that $m_{\sigma}^*$ (``effective mass'' of $\sigma$ 
 defined as a zero of the real part of the propagator
 ${\rm Re}D_{\sigma}^{-1}(\omega = m_{\sigma}^*)=0$)
  approaches to $ m_{\pi}$.  Therefore,
 there exists a density $\rho_c$ at which 
 ${\rm Re} D_{\sigma}^{-1}(\omega = 2m_{\pi})$
 vanishes even before the complete $\sigma$-$\pi$
 degeneracy takes place; namely
 ${\rm Re} D_{\sigma}^{-1} (\omega = 2 m_{\pi}) =
 [\omega^2 - m_{ \sigma}^2 -
 {\rm Re} \Sigma_{\sigma} ]_{\omega = 2 m_{\pi}} = 0$.
At this point, the spectral function is solely dictated by the
 imaginary part of the self-energy;
\beq
\rho_{\sigma} (\omega \simeq  2 m_{\pi}) 
 =  - {1 \over \pi \ {\rm Im}\Sigma_{\sigma} }
 \propto {\theta(\omega - 2 m_{\pi}) 
 \over \sqrt{1-{4m_{\pi}^2 \over \omega^2}}}.
\eeq
 This is a general phenomenon correlated with the 
 partial restoration of chiral symmetry.

We parametrize the chiral condensate in nuclear matter
 $\langle \sigma \rangle$ as
$\langle \sigma \rangle \equiv  \sigma_0 \ \Phi(\rho)$.
In the linear density approximation,
 $\Phi(\rho) = 1 - C \rho / \rho_0 $ with
 $C = (g_{\rm s} /\sigma_0 m_{\sigma}^2) \rho_0$.
 Instead of using $g_{\rm s}$, we  use $\Phi$  as a basic parameter in the 
 following analysis.  The plausible value of $\Phi(\rho = \rho_0)$ is
 0.7 $\sim$ 0.9 \cite{brown}.

\begin{figure}[tbh]
\vspace{8.30cm}
\includegraphics{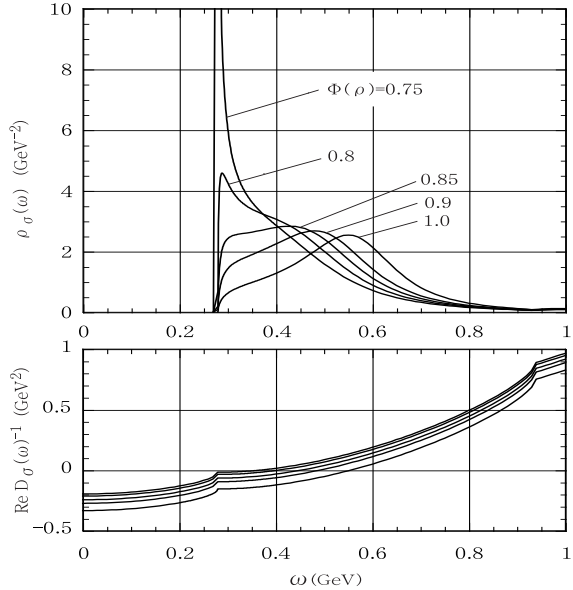}
\caption{Spectral function for $\sigma$ and  the 
 real part of the inverse propagator for several values of
 $\Phi = \la \sigma \ra / \sigma_0$ with
 $m_{\sigma}^{peak} = 550$ MeV. In the lower panel,
 $\Phi$ decreases from bottom to top.}
\label{fig.1}
\end{figure}
 
The spectral function together with ${\rm Re} D_{\sigma}^{-1}(\omega)$  
 calculated with a linear sigma model are shown 
  in Fig.1: The characteristic enhancements of the spectral
 function just above the 2$m_{\pi}$.
The mechanism of the enhancement is understood as follows.
The partial restoration
 of chiral symmetry implies that  $m_{\sigma}^*$ approaches toward
 $m_{\pi}$. On the other hand,
 ${\rm Re}D^{-1}(\omega)$ has a cusp at $\omega = 2 m_{\pi}$.
The cusp point goes up with the density and
eventually hits the real axis at $\rho = \rho_c$
 because ${\rm Re}D^{-1}(\omega )$ increases associated
 with $m_{\sigma}^* \rightarrow 2 m_{\pi}$.
It is also to be noted that even before
the $\sigma$-meson mass $m_{\sigma}^*$ and $m_{\pi}$ in the medium 
are degenerate,i.e., the chiral-restoring point, 
 a large enhancement
 of the spectral function near the $2m_{\pi}$ is seen.

To confirm the threshold enhancement,
measuring 2$\pi^0$ and 
$2\gamma$ in experiments with hadron/photon beams off
 the  heavy nuclear targets are useful. 
 Measuring $\sigma \rightarrow 2 \pi^0 \rightarrow
  4\gamma$ is experimentally feasible 
 \cite{4gamma}, which is free from the $\rho$ meson meson background
  inherent in the $\pi^+\pi^-$ measurement.
 Measuring of 2 $\gamma$'s from the electromagnetic decay of the $\sigma$
 is interesting because of the small final state
 interactions, although the branching ratio is small.\footnote{
One needs also to fight with large 
 background of photons mainly coming from $\pi^0$s.}
 Nevertheless,  if the enhancement is prominent,
  there is a chance to find the signal.  
 When $\sigma$ has a finite three momentum,
one can detect dileptons  
 through the scalar-vector mixing in matter: $\sigma \to \gamma^* \to
 e^+ e^-$. 
We remark that (d, $^3$He)  reactions is also useful to produce 
 the  excitations in the $\sigma$ channel in a nucleus because of the
large incident flux, as  the $\eta$ production\cite{HHG}.
 The incident kinetic energy $E$ of the  deuteron in the laboratory
 system is  estimated to be
  $1.1 {\rm GeV} < E < 10$ GeV, 
  to cover the spectral function 
 in the range  $2m_{\pi} < \omega < 750$ MeV.

Recently  CHAOS collaboration  \cite{chaos} measured the 
$\pi^{+}\pi^{\pm}$
invariant mass distribution $M^A_{\pi^{+}\pi^{\pm}}$ in the
 reaction $A(\pi^+, \pi^{+}\pi^{\pm})X$ with the 
 mass number $A$ ranging
 from 2 to 208: They observed that
the   yield for  $M^A_{\pi^{+}\pi^{-}}$ 
 near the 2$m_{\pi}$ threshold is close to zero 
for $A=2$, but increases dramatically with increasing $A$. They
identified that the $\pi^{+}\pi^{-}$ pairs in this range of
 $M^A_{\pi^{+}\pi^{-}}$ is in the $I=J=0$ state.
The $A$ dependence of the 
 the invariant mass distribution presented in \cite{chaos} 
 near 2$m_{\pi}$ threshold has a close
 resemblance to our model calculation  shown in Fig.1, which suggests
 that this experiment may already provide
  a hint about how the partial restoration of chiral symmetry
 manifest itself at finite density\cite{HKS}.
 
In the present calculation, two-loop diagrams are not included.
Such  diagrams involve the process which gives rise to a change
of the dispersion relation of the pion in the medium.
Indeed, such a change of the dispersion of the pion has
 been proposed as a mechanism to account for the CHAOS data\cite{wambach}.
Clearly, more theoretical and experimental studies are needed to 
make the underlying physics clear of the CHAOS data.

\section{Summary}

In this report, we have emphasized that the $\sigma$ meson
 is a quantum fluctuation of the order parameter of the chiral transition 
 in QCD; the $\sigma$ is analogous to the Higgs particle in the standard
 model.
If the $\sigma$ exists, 
the collective mode in the $I=J=0$ channel as summarized by the $\sigma$ 
  can account for various phenomena in hadron physics which otherwise
  remain mysterious. Thus, it is of fundamental importance to identify 
  the $\sigma$ meson in the free space; if the $\sigma$ is not identified
  experimentally, we must clarify the reason why it is not seen
  experimentally and what is going on in the $\sigma$ channel.
However, evidences have been and are accumulating on
the existence of the $\sigma$ pole in the
scattering matrices in various reactions involving the $\sigma$ channel,
 as seen 
 in this workshop. 

Changes of the environment as characterized by temperature $T$ and/or the 
density $\rho_B$ cause changes of hadron properties, especially those
with collective properties. The $\sigma$ is such an example related with 
the chiral transition in QCD.
We have proposed several experiments to produce the $\sigma$ meson 
using nuclear targets.
We have also given an interpretation about the experimental data
 by CHAOS group: An enhancement of the spectral function in the 
 $\sigma$ channel near 2$m_{\pi}$
 threshold may be a precursor of chiral restoration in the nuclear medium,
  i.e., an evidence of the partial restoration of the chiral symmetry.
We have proposed several experiments to confirm this interpretation.

\vspace{.3cm}
In conclusion, the author thanks T. Hatsuda and H. Shimizu for
their collaboration.

\end{document}